%% file: main.tex
\documentclass[conference]{IEEEtran}
\usepackage{balance}
\usepackage{cite}
\usepackage{url}
\usepackage{multirow}
\usepackage{hyperref}
\usepackage{booktabs}
\usepackage{amsmath,amssymb,amsfonts}
\usepackage{algorithmic}
\usepackage{graphicx}
\usepackage{textcomp}
\usepackage{xcolor}
\usepackage[a4paper, total={184mm,239mm}]{geometry}
\def\BibTeX{{\rm B\kern-.05em{\sc i\kern-.025em b}\kern-.08em
    T\kern-.1667em\lower.7ex\hbox{E}\kern-.125emX}}

\begin{document}

\title{Leveraging Compute-in-Memory for Efficient Generative Model Inference in TPUs}

\author{\IEEEauthorblockN{Zhantong Zhu$^{1,2}$, Hongou Li$^{1,2}$, Wenjie Ren$^{1}$, Meng Wu$^{1}$, Le Ye$^{1}$, Ru Huang$^{1}$ and Tianyu Jia$^{1}$\IEEEauthorrefmark{2}}
\IEEEauthorblockA{$^{1}$School of Integrated Circuits, Peking University, Beijing, China}
\IEEEauthorblockA{$^{2}$School of EECS, Peking University, Beijing, China}
\IEEEauthorblockA{\IEEEauthorrefmark{2}Corresponding Author: tianyuj@pku.edu.cn}
}

\maketitle

\input{0_abstract}

\input{1_introduction}

\input{2_background}

\input{3_architecture_modeling}

\input{4_model_analysis}

\input{5_evaluation_and_discussion}

\input{6_conclusion}

\newpage
\balance
\bibliographystyle{ieeetr}
\bibliography{reference}

\end{document}

%% file: 0_abstract.tex
\begin{abstract}
With the rapid advent of generative models, efficiently deploying these models on specialized hardware has become critical. Tensor Processing Units (TPUs) are designed to accelerate AI workloads, but their high power consumption necessitates innovations for improving efficiency. Compute-in-memory (CIM) has emerged as a promising paradigm with superior area and energy efficiency.  
In this work, we present a TPU architecture that integrates digital CIM to replace conventional digital systolic arrays in matrix multiply units (MXUs). We first establish a CIM-based TPU architecture model and simulator to evaluate the benefits of CIM for diverse generative model inference. Building upon the observed design insights, we further explore various CIM-based TPU architectural design choices. Up to 44.2\% and 33.8\% performance improvement for large language model and diffusion transformer inference, and 27.3$\times$ reduction in MXU energy consumption can be achieved with different design choices, compared to the baseline TPUv4i architecture.
\end{abstract}


%% file: 1_introduction.tex
\section{Introduction}
Generative models, such as large language models (LLMs) and diffusion models (DMs), have exhibited exceptional performance in generating content across various modalities. For example, LLMs have dominated NLP tasks, powering applications like ChatGPT \cite{chatpgt}. DMs have achieved state-of-the-art performance in image and video generation, e.g. OpenAI's DALL-E 3, Stability AI's Stable Diffusion (SD) \cite{esser_scaling_2024}, and OpenSORA \cite{opensora}. The growing demand for AI generative models underscores the necessity of designing high-performance acceleration hardware for the model deployment. Currently, GPUs and Tensor Processing Units (TPUs) serve as the primary hardware platforms for AI inference and training, featuring massive parallel computing components, e.g., Tensor Cores in NVIDIA GPUs \cite{choquette_nvidia_2021} and matrix multiply units (MXUs) in Google TPUs \cite{tpuv4i_2021, tpuv4_2023}. However, these mainstream acceleration hardware typically consumes over 350W of TDP power, necessitating cross-stack innovations to enhance their computational efficiency.

\begin{figure}[t]
  \centering
  \includegraphics[width=\linewidth]{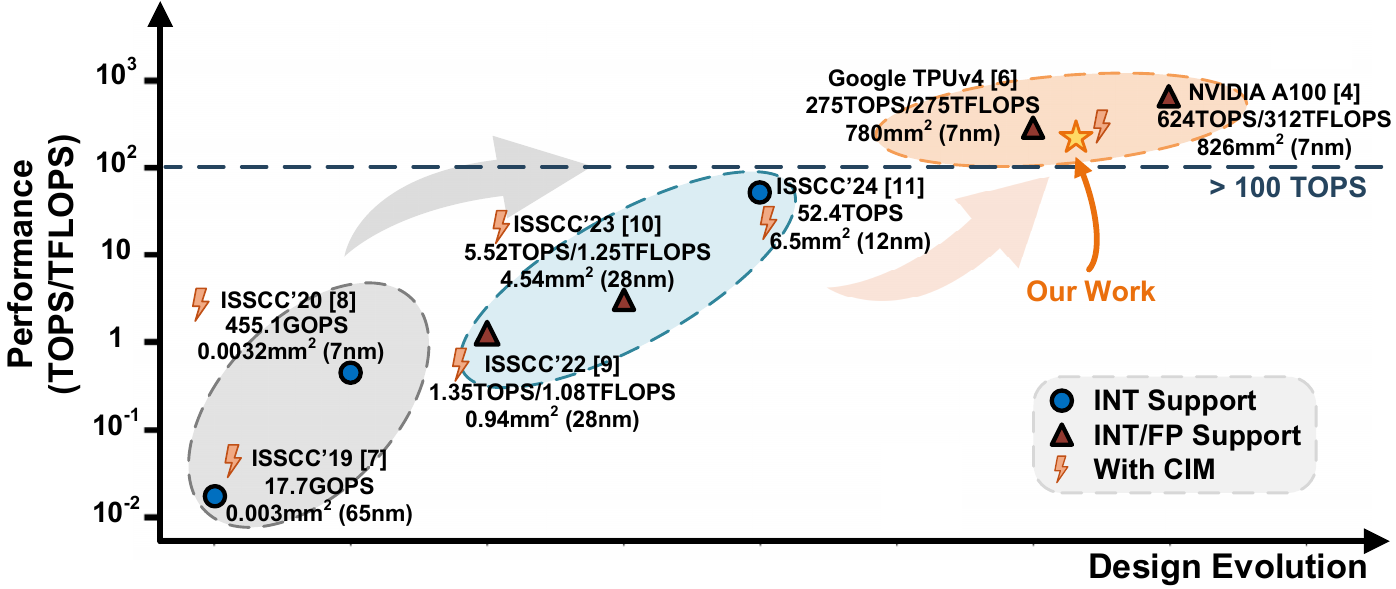}
  \caption{Evolution of the computing performance of CIM-based designs.}
  \label{fig: cim_design}
  \vspace{-10pt}
\end{figure}

In recent years, compute-in-memory (CIM) technique has emerged as a promising design approach, offering impressive energy efficiency and computational density. Fig. \ref{fig: cim_design} illustrates the performance evolution of CIM-based designs over the past few years. Early efforts in CIM macros achieved performance levels below 500GOPS with high efficiency \cite{si_twin-8t_2019, dong_153_2020}.
Subsequent developments incorporated multiple CIM macros into a larger core-level design, achieving over 1TOPS performance \cite{tu_28nm_2022, yue_28nm_2023, hager_metis_2024}. Recently, CIM-based AI chip product have also been developed, such as a quad-core SoC from Axelera AI that delivers 52.4TOPS per core \cite{hager_metis_2024}, demonstrating the scalability of CIM for high-performance chips. Despite these advancements, there is still a significant performance gap between CIM-based AI chips and established accelerators like GPUs or TPUs. For instance, the NVIDIA A100 GPU achieves 312TFLOPS@BF16 \cite{choquette_nvidia_2021}, while the Google TPUv4 delivers 275TFLOPS@BF16 \cite{tpuv4_2023}. 

It is a natural thought to utilize CIM technique to enhance the efficiency of TPUs or GPUs. In this work, we explore the design methodology and benefits for CIM-based TPUs.
Two key questions are explored by this paper.
\textit{First}, what efficiency improvements can CIM bring to TPUs?
\textit{Second}, how can we redesign CIM-based TPUs to optimize performance for various generative models?
To explore these questions, we begin by establishing a comprehensive architecture modeling for CIM-based TPUs, building on the TPUv4i architecture \cite{tpuv4i_2021}. We then analyze the computational characteristics of diverse generative models, including mainstream LLMs and DMs, on our CIM-based TPU model. It is observed that significant area and energy benefits can be obtained by leveraging CIM technique for TPU. 

Based on the observed design insights, we further explore the optimizations of existing TPU architecture with different design choices of CIM-based MXUs. Compared to the baseline TPUv4i, a maximum 44.2\% and 33.8\% performance improvement can be obtained for LLM and DM inference, respectively. Moreover, a maximum 27.3$\times$ reduction in MXU energy consumption can be obtained by leveraging CIM technique. To the best of our knowledge, this is the first architectural modeling, analysis, and design exploration of a CIM-based TPU. Our experiments and observations will provide valuable insights for designing the next generation of efficient, high-performance acceleration hardware.

The contributions of this work are summarized as follows.
\begin{itemize}
    \item We establish an architecture model and simulator to evaluate design benefits of CIM-based TPUs.
    \item We analyze the inference characteristics and CIM benefits for generative models on CIM-based TPUs. 
    \item We further explore architecture optimizations for CIM-based TPUs, with substantial improvement obtained.
\end{itemize}

%% file: 2_background.tex
\section{Background}

\subsection{Generative AI Models}
\begin{figure}[t]
  \centering
  \includegraphics[width=0.96\linewidth]{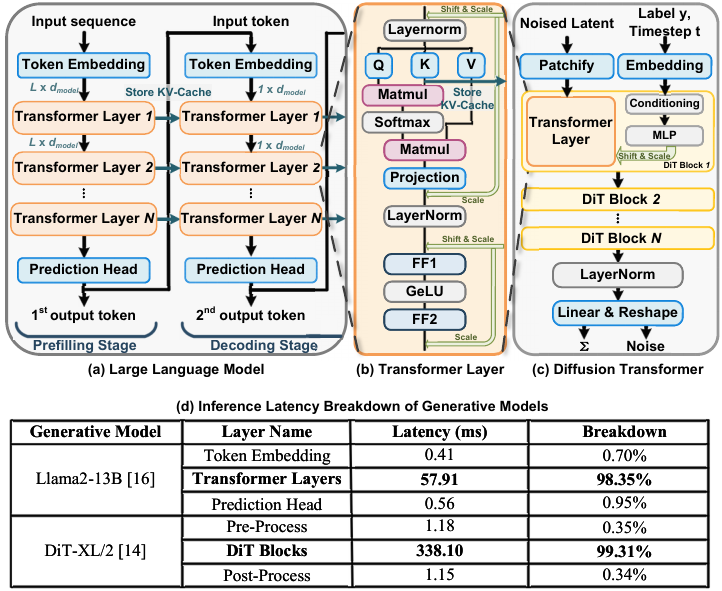}
  \caption{Generative model architecture and runtime breakdown.}
  \label{fig: model_breakdown}
  \vspace{-10pt}
\end{figure}

Among generative models, two prominent model architectures are LLMs and DMs, as illustrated in Fig. \ref{fig: model_breakdown}.
The inference of LLMs is primarily driven by a stack of Transformer layers. Before these, a token embedding step processes the input prompt sequence. A prediction head generates the next output token after the Transformer layers.
LLMs inference consists of two distinct stages: \textit{Prefilling} (Summarization) and \textit{Decoding} (Generation).
During Prefilling stage, the model takes a prompt sequence as input and generates a Key-Value Cache (KV Cache) for each Transformer layer.
In the Decoding stage, the model iteratively takes one token as input, generates the next token, and updates the KV Cache by incorporating the results from the current iteration.
Notably, these two stages exhibit distinct characteristics, i.e., Prefilling is primarily compute-bound and Decoding is memory-bound \cite{yuan_llm_2024}.
Since the capabilities of LLMs are closely tied to model size \cite{zhao_survey_2023}, the unprecedented scaling of model size exacerbates the challenges of designing efficient hardware for LLM deployment.

Diffusion Transformer (DiT) \cite{peebles_scalable_2023} is a new class of DMs based on the Transformer architecture instead of the U-Net architecture adopted in earlier DMs \cite{rombach_high-resolution_2022}. The DiT architecture has demonstrated strong performance and has been widely adopted in recent DMs, such as SD v3 \cite{esser_scaling_2024}.
As illustrated in Fig. \ref{fig: model_breakdown} (c), the core component of DiT inference process is the sequence of DiT blocks. Each block includes a Transformer layer, augmented with conditioning, shift and scale operations.

It is clear that the core architecture of LLMs and DiTs is the Transformer architecture. 
We conducted inference evaluations to analyze the runtime breakdown of these generative models on NVIDIA A100 GPUs \footnote{DiT inference was run on a single A100-PCIe-40GB GPU, while Llama2-13B inference utilized two GPUs.}.
Analysis targets included Llama2-13B \cite{touvron_2023_llama2} with the Alpaca dataset \cite{alpaca} and DiT-XL/2 with an image resolution of 512$\times$512, as shown in Fig. \ref{fig: model_breakdown} (d).
It is observed that the Transformer layers dominate the inference time, accounting for 98.35\% in Llama2-13B and 99.31\% in DiT-XL/2. 
By contrast, other components contribute minimally to the overall inference time. For Llama2-13B, the token embedding and prediction head account for only 0.70\% and 0.95\%, respectively.
Similarly, in DiT-XL/2, the pre-processing (patchify and embedding) and post-processing (LayerNorm, linear \& reshape) layers only account for 0.35\% and 0.34\% of the total inference time.

\subsection{Compute-in-Memory for Efficient Computing}
Compute-in-memory is a efficient computing paradigm to alleviate the efficiency challenge by fusing MAC operations into memory.
Early CIM designs utilize analog-domain computation \cite{liu_fully_2020}, performing MAC operations in voltage, charge or time domain to achieve extremely high energy efficiency.
However, analog CIMs face limitations for large-scale implementation due to notable non-idealities such as process variations and low bit precision \cite{yu_compute--memory_2021}.
Therefore, we focus on digital SRAM-based CIM using digital-domain computation \cite{hager_metis_2024}, which offers greater robustness and flexibility for scalable, high-performance hardware design.

A typical digital CIM macro is organized into multiple banks, each corresponding to one output channel.
Within each bank, the bitcell array is further divided into sub-arrays, with each sub-array handling one input channel \cite{si_twin-8t_2019, dong_153_2020}.
For example, \cite{dong_153_2020} presents a 7nm CIM macro with an average energy efficiency of 351TOPS/W@INT4.
Beyond INT operations, recent digital CIM designs have also incorporated floating-point support for both training and inference while preserving high energy and area efficiency \cite{tu_28nm_2022, yue_28nm_2023, guo_28nm_2023}.
For instance, \cite{guo_28nm_2023} showcases an SRAM-based CIM macro using a cell array with multi-precision floating-point computing units, achieving SOTA energy efficiency of 31.6TFLOPS/W and area efficiency of 2.05TFLOPS/mm$^2$.
CIM macros have already been adopted into AI processor designs \cite{tu_28nm_2022, yue_28nm_2023, jing_nerf_2024}.
For instance, \cite{tu_28nm_2022} demonstrates a digital CIM processor capable of INT8/INT16/BF16/FP32 operations through a unified FP/INT pipeline, achieving 36.5TOPS/W@INT8 and 29.2TFLOPS/W@BF16 system efficiency.

By contrast, TPUs with advanced technology nodes exhibit over 10$\times$ lower efficiency. For instance, TPUv4i delivers a peak performance of 138TFLOPS@BF16 with 175W at 7nm \cite{tpuv4i_2021}, resulting in an efficiency of 0.788TFLOPS/W@BF16. This is an order of magnitude lower than the efficiency of CIM-based designs, e.g., 15TOPS/W efficiency demonstrated by a CIM SoC \cite{hager_metis_2024}.

%% file: 3_architecture_modeling.tex
\section{Architecture Modeling for CIM-based TPUs}
\label{sec: arch_modeling}
\subsection{Architecture Overview}

\begin{figure}[t]
  \centering
  \includegraphics[width=0.9\linewidth]{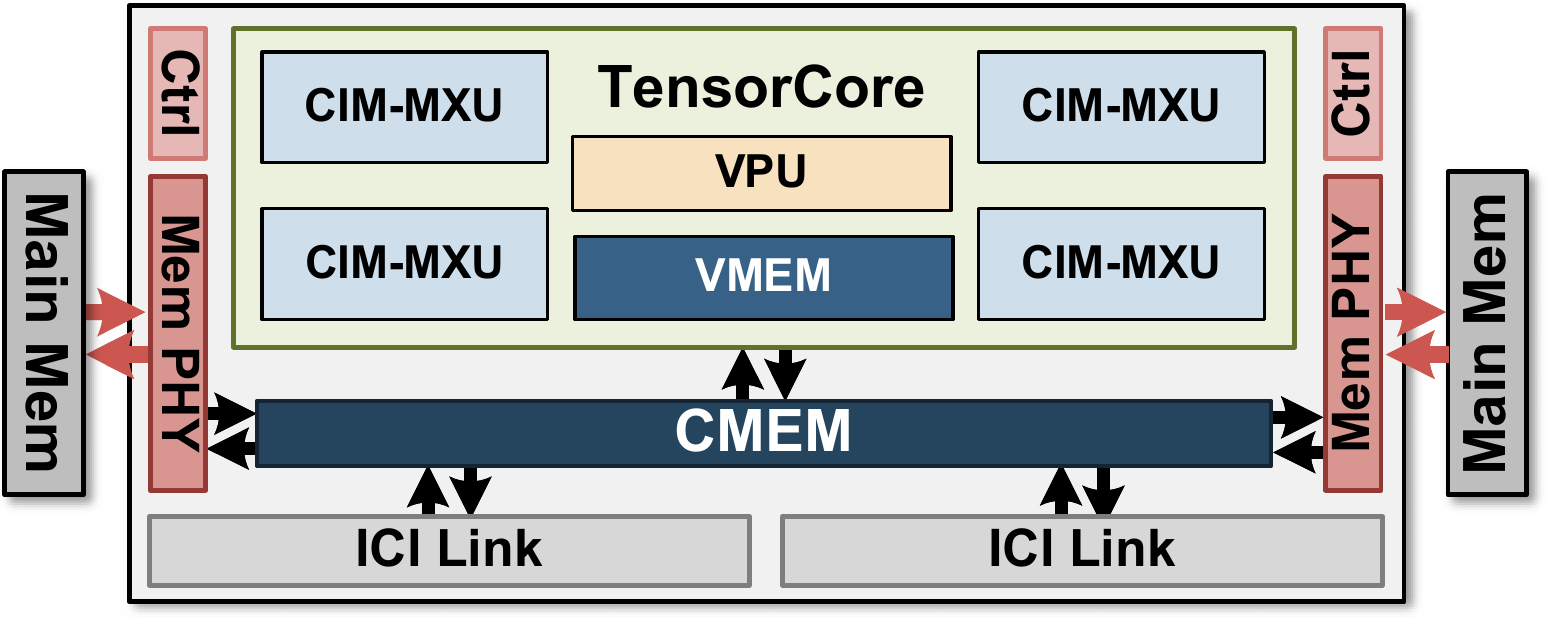}
  \caption{Architecture modeling of CIM-based TPU.}
  \label{fig: hardware_template}
  \vspace{-10pt}
\end{figure}

\begin{table}[t]
    \caption{Architecture parameters for CIM-based TPU.}
    \begin{center}
        \begin{tabular}{|c|c|c|}
            \hline
            \textbf{Key parameters} & \textbf{TPUv4i \cite{tpuv4i_2021}} & \textbf{CIM-based TPU} \\
            \hline
            Tensor Core count & 1 & 1 \\
            \hline
            MXU dimension & 128$\times$128 MACs & 16$\times$8 CIMs \\
            \hline
            CIM core dimension & N/A & 128 $\times$ 256 \\
            \hline
            Vector width & \multicolumn{2}{c|}{8 $\times$ 128 } \\
            \hline
            Vector memory size & \multicolumn{2}{c|}{ 16 MB }  \\
            \hline
            Common memory size & \multicolumn{2}{c|}{ 128 MB } \\
            \hline
            Main memory size & \multicolumn{2}{c|}{8 GB } \\
            \hline
            Main memory bandwidth &  \multicolumn{2}{c|}{ 614 GB/s }\\
            \hline
            ICI link bandwidth & \multicolumn{2}{c|}{100 GB/s} \\
            \hline
        \end{tabular}
        \label{tab: example_hardware_parameters}
    \end{center}
    \vspace{-15pt}
\end{table}

In this section, we introduce the architectural modeling of our CIM-based TPU, designed for generative model inference. 
Fig. \ref{fig: hardware_template} illustrates the architecture, which is based on the TPUv4i inference-only accelerator \cite{tpuv4i_2021}.
The TPUv4i performs AI computations using a TensorCore, which consists of four matrix multiply units (MXUs) for matrix multiplication, a vector processing unit (VPU) and vector memory (VMEM).
The original MXUs are implemented as 128$\times$128 systolic arrays of Multiply-Accumulate (MAC) units for parallel computation.
In our model, we replace the vanilla MXU with our CIM-based MXU (CIM-MXU, details in Sec. \ref{para: cim_mxu}) to enhance matrix computation efficiency. 
The VPU remains unchanged to process other general parallel operators, such as normalization and activation functions.
At the top level, the chip includes 128MB of common memory (CMEM), an on-chip interconnect (OCI), two chip-to-chip interconnect (ICI) links, and interfaces to main memory, e.g. HBMs.
Both CMEM and VMEM are implemented as on-chip SRAM.
We model the memory access between VMEM and CMEM using available OCI bandwidth.

To ensure accurate performance evaluation, our model leverages the prior architecture template from \cite{zhang_hardware_2023}, which has been validated for accurate LLM inference, and incorporates an accurate CIM model for our CIM-based TPU.
Unlike prior CIM simulator \cite{andrulis_cimloop_2024}, our approach maintains the two-level memory hierarchy as used in TPUs. 
Key architectural parameters, such as CIM-MXU counts and dimensions, buffer sizes, and OCI/ICI bandwidth, are fully configurable, as outlined in Table \ref{tab: example_hardware_parameters}.
With minor modifications, our architecture modeling can also be adapted to other TPU variants or GPUs.

\subsection{Design and Modeling of CIM-MXU}
\label{para: cim_mxu}

\begin{figure}[t]
  \centering
  \includegraphics[width=\linewidth]{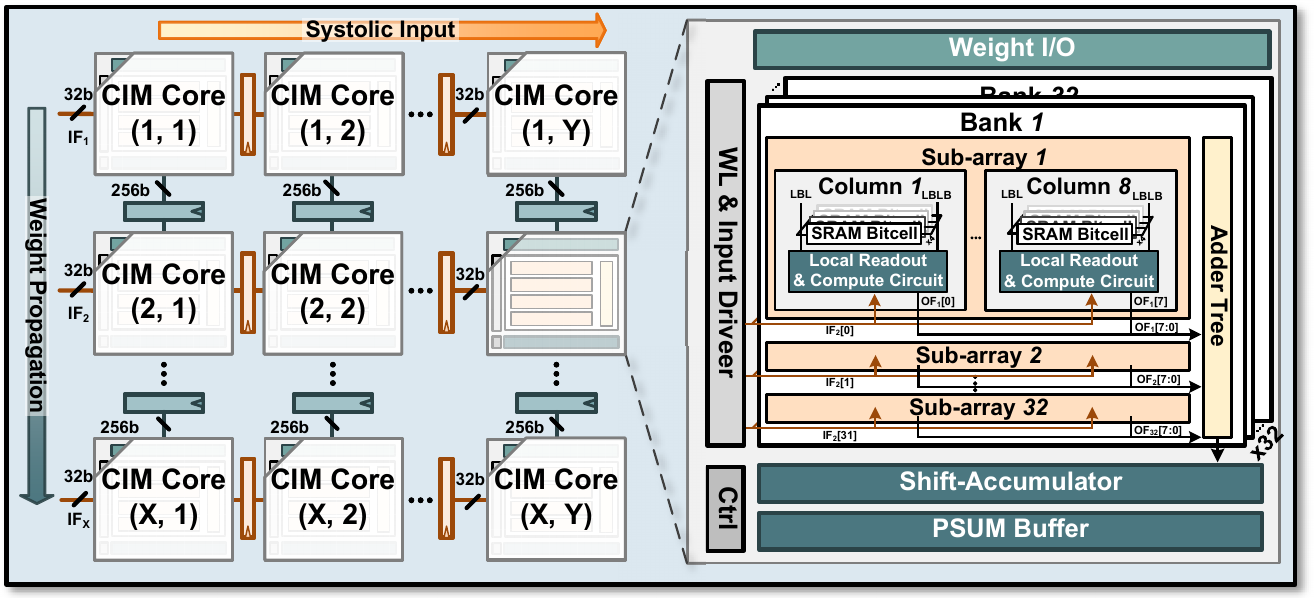}
  \caption{Architecture and CIM design details of CIM-MXU.}
  \label{fig: cim_mxu_arch}
  \vspace{-10 pt}
\end{figure}

CIM-MXU is designed to replace conventional MXU in TPUs to accelerate GEMM/GEMV operations.
However, two challenges arise when organizing CIM macros for the high parallelism in a systolic array.
First, each CIM macro has limited dimensions, making it difficult to scale up or scale out.
Most CIM designs feature bespoke circuit design, meaning that modifying the data path to increase macro size (\textit{scaling up}) is challenging.
Meanwhile, the bit width of input activation and weight will increase significantly when organizing multiple CIM macros (\textit{scaling out}) for simultaneous computation.
Second, systolic matrix multiplications require frequent weight updates to maintain high hardware utilization.
Furthermore, GEMM/GEMV operations in Transformers often have very low weight reuse rate, leading to frequent weight matrix updating.

In our work, we adopt a systolic data path to incorporate multiple CIM macros within the CIM-MXU, as shown in Fig. \ref{fig: cim_mxu_arch}.
At the top level, a 16$\times$8 grid of CIM cores forms a two-dimensional systolic array, where 128 MAC operations are performed each cycle within each CIM core.
The CIM computation aligns with typical weight-stationary digital CIMs \cite{guo_28nm_2023}, where the input vector broadcast to all output channels in a bit-serial manner.
An output-stationary dataflow is employed in the CIM-MXU systolic array, with inputs and weights being propagated after each computation wave.
In the row dimension, a 32-bit input vector is propagated systolically, moving sequentially across the CIM core columns.
In the column dimension, each CIM core can perform CIM read/write operations through its dedicated weight I/O.
During computation, weight matrices are propagated to the CIM core in adjacent row via interleaved SRAM read/write operations.
To accommodate frequent weight updates, our CIM macro supports simultaneous computation and weight read/write operations via the weight I/O, which is similar as \cite{10067555}.
Such systolic data path maximizes weight and input reuse, conserving IO bandwidth and enabling the system to scale out to larger CIM core arrays.

Our CIM-MXU can perform both BF16 and INT8 operations as the original MXU in TPUv4i. 
In FP mode, mantissa bits of the weight matrix are loaded into CIM macros, and the input activation is processed by a pre-processing unit before the mantissa bits are transferred to the CIM array. The pre-processing unit performs exponent alignment and mantissa shifting for the INT MAC in CIM macro, and a post-processing unit handles the rest shift-and-accumulation and rounding operations.
In INT mode, the pre-processing unit is bypassed, allowing direct loading of input activation into the CIM array.

\subsection{Workload Evaluations}
\label{para: perf_eval}

\begin{figure}[t]
  \centering
  \includegraphics[width=0.9\linewidth]{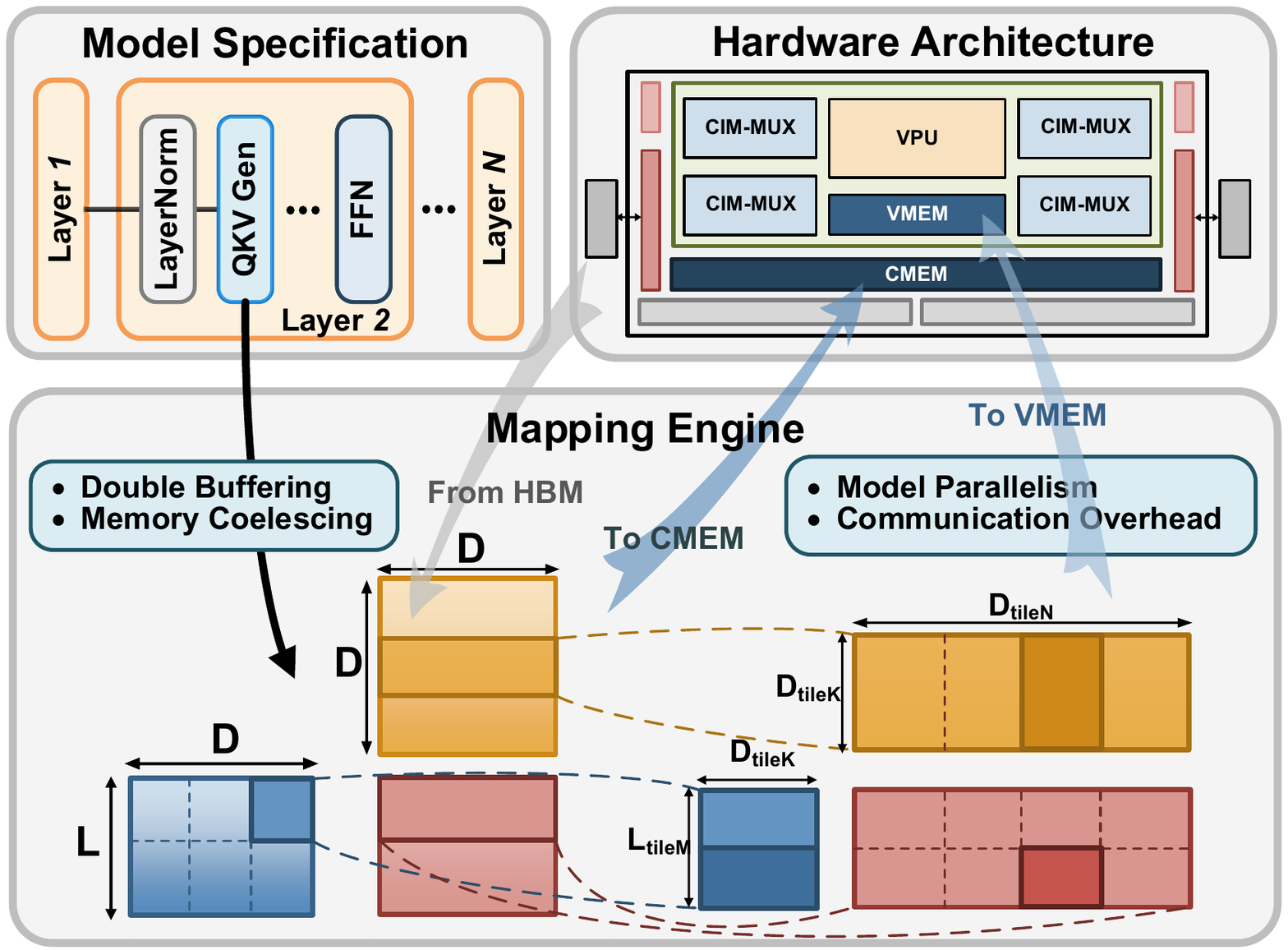}
  \caption{Workload evaluations with a mapping engine for CIM-based TPUs.}
  \label{fig: simulator}
  \vspace{-10pt}
\end{figure}

\textbf{Mapping and scheduling.}
Given the computational graph and hardware configurations, a model mapping engine performs tiling and scheduling of operators onto CIM-based TPU.
Fig. \ref{fig: simulator} provides a mapping example for a single layer.
The model with a size of $[L, D] \times [D, D] = [L, D]$ will be partitioned into subtiles with size of $[L_{tile_M}, D_{tile_K}] \times [D_{tile_K}, D_{tile_N}]$ to fit into the on-chip CMEM. 
The tensors will be further partitioned to fit into the VMEM before being processed by CIM-MXUs or VPU.
Since the objective space of valid mappings is very large, we prune the mapspace using heuristics similar as prior work \cite{parashar_timeloop_2019, zhang_hardware_2023}.
The mapping engine further explores the performance-optimal mapping to better utilize hardware resources.
To overlap computation with memory access cycles, we utilize double buffering and memory coalesce technique at each level of the memory hierarchy as scheduling options.

\textbf{Computation evaluation.}
Our CIM-based TPU supports the evaluations of a wide range of key operators in generative models, including both GEMM/GEMV and other non-linear functions.
GEMM operates on three-dimensional data, in which the input matrices are tiled into smaller matrices to fit into SRAM buffers.
For the baseline comparison, we use SCALE-Sim \cite{samajdar_systematic_2020} to evaluate systolic arrays with a given array dimension, along with input and weight matrix sizes.
We also model the computation of Softmax, LayerNorm and GeLU with similar methodology. These operators utilize vector units instead of CIM cores. We implement Softmax with algorithm \cite{milakov_online_2018} and approximate GeLU with \textit{tanh}, which is the same approach used in DiTs \cite{peebles_scalable_2023}.
In our performance evaluations, we exploit tensor parallelism and pipeline parallelism \cite{shoeybi_megatron-lm_2020} to scale the computation capability of TPUs. 

%% file: 4_model_analysis.tex
\section{Analysis of Generative Models}
\label{sec: model_analysis}

\subsection{MXU Evaluations}

To validate the CIM benefits, we first demonstrate a comparison between standalone digital MXU and a 8$\times$16 CIM-MXU, as the parameters listed shown in Table \ref{tab: example_hardware_parameters}.
We use Gemmini \cite{gemmini-dac} to generate a 128$\times$128 systolic array, which is physically implemented using Cadence Genus and Innovus to obtain post-P\&R power and area.
The CIM core's layout is manually drawn and the CIM-MXU is implemented using RTL for physical design.
Both digital MXU and CIM-MXU are implemented using the same TSMC 22nm technology.
As shown in Table \ref{tab: cim_and_systolic_array_comparison}, our CIM-MXU implementation has 7.26TOPS/W and 1.31TOPS/mm$^2$ energy and area efficiency, which is 9.43$\times$ and 2.02$\times$ better than digital MXU while maintaining the same MACs per cycle throughput.

\begin{table}[t]
    \caption{Comparison between CIM-MXU and digital MXU.}
    \begin{center}
        \begin{tabular}{|c|c|c|c|}
            \hline
            \textbf{Evaluation Metrics} & \textbf{Digital MXU}  & \textbf{CIM-MXU} & \textbf{Speedup} \\
            \hline
            MACs per cycle & 16384 & 16384 & 1$\times$ \\
            \hline
            Energy Efficiency & 0.77TOPS/W & 7.26TOPS/W & 9.43$\times$ \\
            \hline
            Area Efficiency & 0.648TOPS/mm$^2$ & 1.31TOPS/mm$^2 $ & 2.02$\times$ \\
            \hline
        \end{tabular}
        \label{tab: cim_and_systolic_array_comparison}
    \end{center}
    \vspace{-10pt}
\end{table}

\subsection{Model Inference Evaluations}

\begin{figure}[t]
  \centering
  \includegraphics[width=\linewidth]{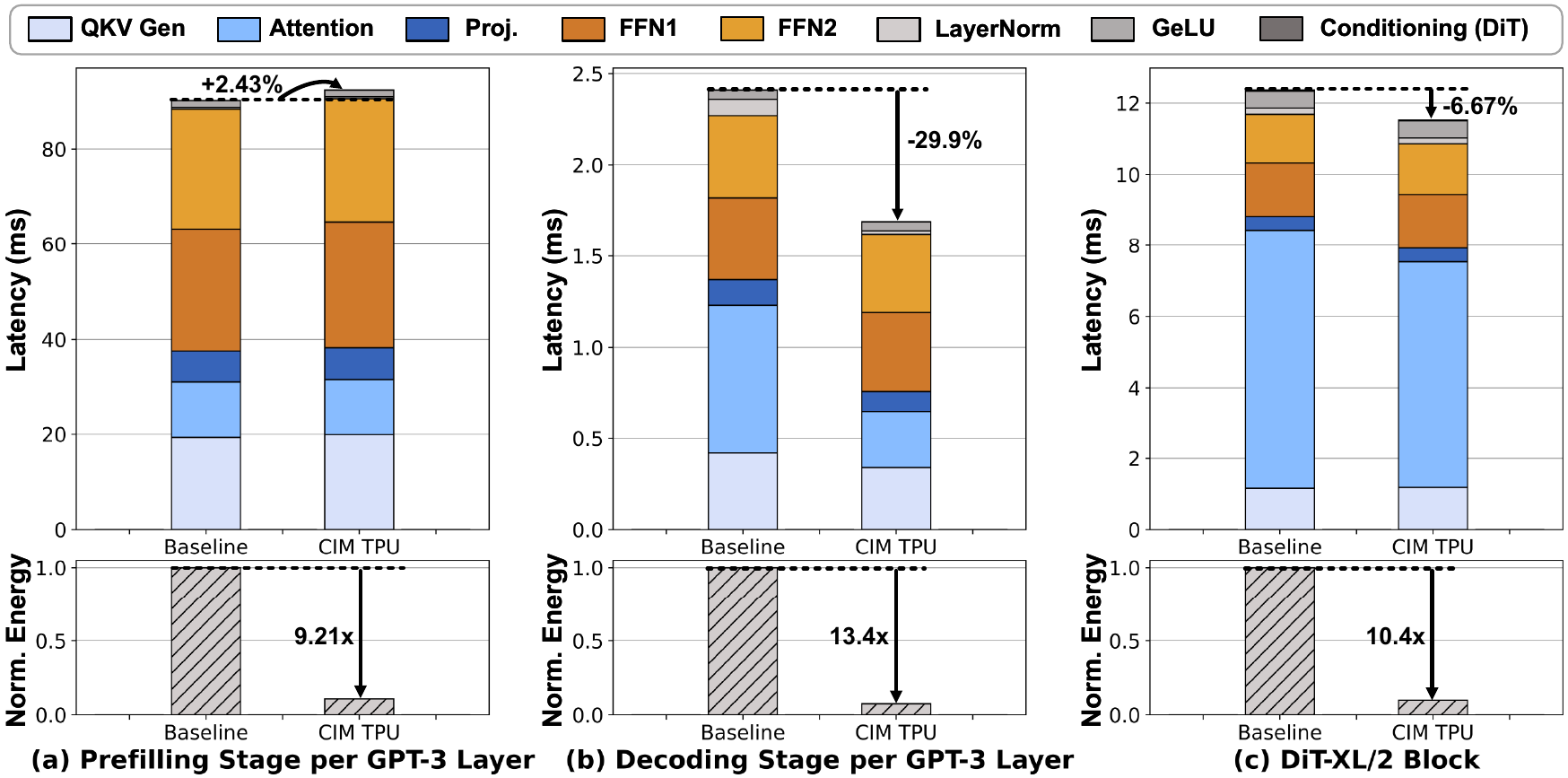}
  \vspace{-20pt}
  \caption{Comparison between baseline and CIM-based TPU designs.}
  \label{fig: per_layer_comparison}
  \vspace{-15pt}
\end{figure}

To further assess the effectiveness of model inference on TPUs, we select two representative generative models for evaluations, i.e., GPT-3 \cite{brown_language_2020} and DiT-XL/2.
We adopt the original TPUv4i architecture parameters as comparison baseline and our CIM-based TPU replaces MXUs while maintaining other hardware specifications, such as memory capacity and bandwidth.
Both baseline TPU and CIM-based TPU are scaled to the same technology and frequency for fair performance and energy comparisons.
The model configurations of Transformer layers in target generative models are listed in Table \ref{tab: model_config}.
We set batch size to 8 and simulate the inference of a single Transformer layer from GPT-3, and one DiT block from DiT-XL/2 with image resolution of 512$\times$512, using INT8 data precision.
During Prefilling stage, we set the input token length to 1024. For Decoding stage, we simulate the processing of the 256$^{th}$ output token.
Fig. \ref{fig: per_layer_comparison} shows the generative model inference latency and energy consumption of MXUs for the evaluated models.

\begin{table}[t]
    \caption{Configurations of evaluated generative models}
    \begin{center}
        \begin{tabular}{|c|c|c|c|}
            \hline
            \textbf{Generative model} & \textbf{\# Layers} & \textbf{\# Heads} & $ {\textbf{\textit{d}}}_{\textbf{\textit{model}}} $ \\
            \hline
            GPT3-30B & 48 & 56 & 7168 \\
            \hline
            DiT-XL/2 & 28 & 16 & 1152\\
            \hline
        \end{tabular}
        \label{tab: model_config}
    \end{center}
    \vspace{-10pt}
\end{table}

\textbf{LLM Prefilling:} During Prefilling stage, QKV generation, Projection, and FFN layers consist of GEMM operations with large matrix dimensions. Hence it is observed that these layers take up 84.9\% of TPU inference latency, marking them as the primary computation bottleneck.
In contrast, Attention layers, which consists of \textit{Q$\times$K$^T$}, \textit{S$\times$V$^T$}, and Softmax operations, contribute only 13.1\% to overall latency.
Since the systolic array in baseline MXUs has already been optimized for large GEMM operations, our CIM-MXU will be not bring inference latency improvement.
However, the energy efficiency advantage of CIM-MXU leads to 9.21$\times$ less energy consumption than digital MXUs for Prefilling stage. 

\textbf{LLM Decoding:} 
Since the input token length for LLM decoding is one, this significantly reduces the input dimensions of GEMM as GEMV operations, resulting in lower arithmetic intensity and memory bandwidth bottleneck.
In baseline TPU design, Attention layers account for 33.7\% inference latency, primarily driven by \textit{Q$\times$K$^T$} and \textit{S$\times$V$^T$} layers. 
Compared to the baseline design, CIM TPU accelerates these GEMV layers by 72.7\%, leading to a notable 29.9\% inference latency reduction.
This speedup is attributed to the CIM architecture, where the input activation vector in each CIM core is broadcast to all output channels in a bit-serial manner. Such dataflow eliminates the necessary of traversing all preceding MAC units, which is required in conventional systolic array.
Benefiting from both latency and efficiency improvements, CIM-MXUs consume 13.4$\times$ less energy than digital MXU, significantly boosting the LLM decoding efficiency.

\textbf{DiT Block:} 
For a DiT block, the GEMM operations from QKV generation, Projection, and FFN consume up to 35.65\% inference latency.
For these GEMM computations within DiT blocks, both the digital MXU and CIM-MXU exhibits similar performance.
Meanwhile, Softmax computation in Attention layers take up to 36.9\% inference latency, becoming the computation bottleneck in DiT inference.
It is observed that the CIM-MXU has 30.3\% improvement for \textit{Q$\times$K$^T$} and \textit{S$\times$V$^T$} processing inside Attention layers due to better DiT mapping than digital MXU.
Overall, the CIM-based TPU achieves a 6.67\% latency and 10.4$\times$ energy reduction compared to the baseline design.

Based on the above model inference evaluations, we conclude the design observations for adopting CIM in TPUs.

\textbf{CIM can significantly improve area and energy efficiency.}
It is clear that the CIM has notable efficiency advantages for TPUs. 
Our CIM-MXU contains 128 CIM cores, delivering the same peak performance as the baseline MXU with only 50\% area. 
Meanwhile, CIM-MXU can reduce energy consumption by about one order of magnitude, significantly enhancing the energy efficiency of matrix computations.

\textbf{CIM contributes differently for diverse generative models.}
It is observed that CIM gains the most performance benefits for GEMV-dominant layers, such as LLM Decoding. In fact, decoding consumes the most latency for LLMs, due to the large number of output tokens. Hence CIM-based TPU provides valuable performance and efficiency gains for LLM inferences.
In DiT inference, the total inference time involves iteratively processing multiple DiT blocks, where the primary contributors to overall latency are Softmax and GEMM operations.
Hence CIM mainly contributes to area and energy efficiency improvement rather than performance.

%% file: 5_evaluation_and_discussion.tex
\section{Architecture Exploration and Evaluation}
\subsection{Architecture Exploration}
As our prior evaluations indicated, CIM technique provides area and energy savings compared to the baseline design.
This motivates us to capitalize on these power savings and area headroom to further explore CIM-based TPU architecture for optimal design choices.
We present several architecture design options in Table \ref{tab: arch_exploration} and conduct model inference analysis as before.

\begin{table}[t]
    \caption{Architecture design choices of CIM-MXU.}
    \begin{center}
        \begin{tabular}{|c|c|c|c|}
            \hline
            \textbf{Parameters} & \multicolumn{3}{c|}{\textbf{Architecture Choices}} \\
            \hline
            Array dimension & $8 \times 8$ & $16 \times 8$ & 16 $\times$ 16 \\
            \hline
            CIM-MXU count & 2 & 4 & 8 \\
            \hline
        \end{tabular}
        \label{tab: arch_exploration}
    \end{center}
    \vspace{-15pt}
\end{table}

\begin{figure*}[t]
  \centering
  \includegraphics[width=0.92\linewidth]{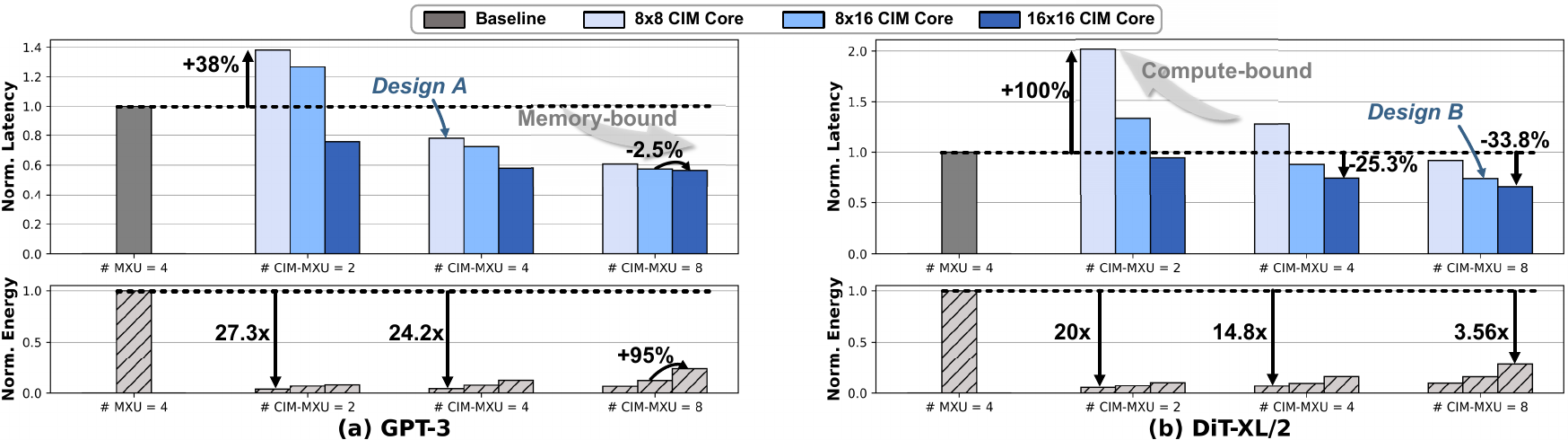}
  \vspace{-5pt}
  \caption{Architecture exploration of different CIM-MXU configurations.}
  \label{fig: arch_explore}
  \vspace{-10pt}
\end{figure*}

Fig. \ref{fig: arch_explore} illustrates the evaluation of GPT-3-30B and DiT-XL/2 inference latency and energy across various CIM-MXU architecture settings compared to the baseline design.

\textbf{LLM Inference.}
Our simulation encompasses both Prefilling and Decoding stages.
We set the input and output sequence lengths to 1024 and 512, respectively, to reflect typical real-world scenarios, in which Decoding dominates the latency and energy consumption of MXUs.
Due to the memory-bound nature of LLM Decoding, CIM-MXUs exhibit limited performance gains when MXU counts and array dimensions continue to increase.
For example, although the 8 CIM-MXU configuration with 16$\times$16 CIM cores has 2$\times$ peak performance compared to the same number of CIM-MXUs with 16$\times$8 CIM cores, only 2.5\% performance improvement is achieved for LLM inference, at the cost of a 95\% energy increase. 
To effectively harness the efficiency advantages of CIM technique, adopting smaller-sized CIM-MXUs can yield significant energy savings with minimal performance degradation.
For example, even with only two CIM-MXUs in the TPU, featuring a smaller 8$\times$8 CIM core array has 38\% latency increase while gaining 27.3$\times$ energy savings.
Considering the trade-off between latency and energy, we adopt four CIM-MXUs with 8$\times$8 array dimension as the optimized architecture for LLM inference, denoted as \textit{Design A}.

\textbf{DiT Inference.} For compute-bound DiT inference, we observe that CIM-based TPUs with more or larger CIM-MXUs achieve better inference latency by leveraging higher peak performance.
Specifically, CIM-based TPUs with 4 and 8 CIM-MXUs (16$\times$16 CIM cores) achieve 25.3\% and 33.8\% inference latency reduction, respectively.
However, this enhanced performance also comes with increased energy cost, since more matrix computation components are added to the TPU.
Fortunately, the high efficiency of CIM ensures that the CIM-MXU configuration with the highest performance, i.e., 8 CIM-MXUs each with 16$\times$16 CIM cores, still consumes 3.56$\times$ less power compared to the vanilla systolic MXU in TPUs.
When MXU performance decreases, DiT inference latency increases linearly, e.g. two CIM-MXUs with 8$\times$8 CIM cores has a 100\% higher latency than the baseline design.
However, while inference runtime becomes longer, the CIM-MXU power is reduced by $20\times$ compared to the baseline due to fewer CIM cores.
Considering latency, energy and area trade-offs of MXUs, we derive an optimal CIM-MXU architecture for DiT inference, featuring 8 CIM-MXUs each with a 16$\times$8 array dimension, which is denoted as \textit{Design B}.

It is observed that none of the optimized TPU designs are ideal for all generative model inferences.
Despite \textit{Design A} having only half the peak performance of the baseline MXU, it allows for more flexible mapping strategies and a higher utilization rate, thereby improving hardware efficiency.
In contrast, \textit{Design B} is equipped with higher performance for faster DiT inference performance with lower energy cost.

\subsection{Evaluation of Multi-Device Inference}

Beyond standalone CIM-based TPU architecture explorations, we also extend our evaluation to multi-TPU inference scenarios, satisfying the need for large-scale deployment of generative models.
To accommodate large batch sizes for model inferences, we scale the number of TPUs and implement up to 4-way pipeline parallelism with 4 TPUs interconnected in a ring topology to fully utilize the two ICI links on each TPU chip, as the default configuration in TPUv4i \cite{tpuv4i_2021}.
We evaluate the inference throughput of generative models for the baseline TPU and our optimized CIM TPUs, \textit{Design A} and \textit{Design B}.

\begin{figure}[t]
  \centering
  \includegraphics[width=\linewidth]{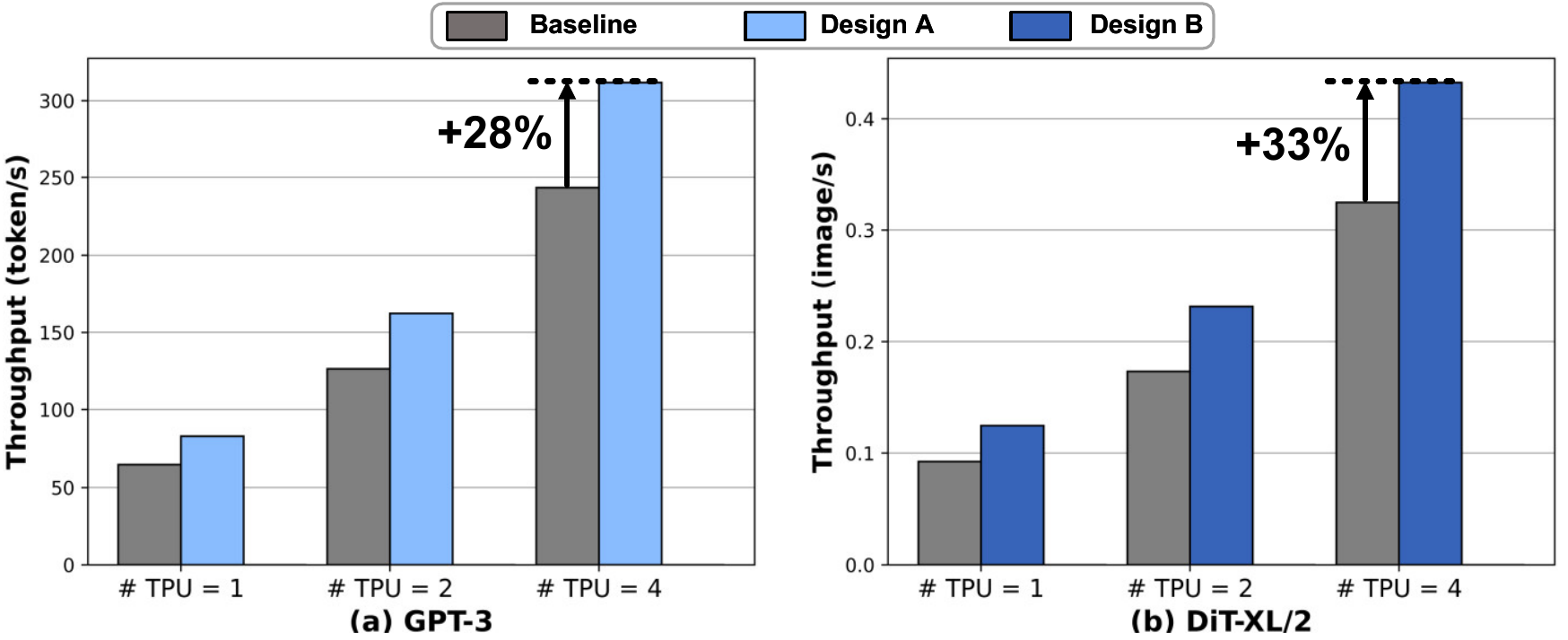}
  \vspace{-15pt}
  \caption{Comparison between inference throughput.}
  \label{fig: multi_device_comp}
  \vspace{-15pt}
\end{figure}

Fig. \ref{fig: multi_device_comp} illustrates the inference throughput of GPT-3-30B and DiT-XL/2 when one, two, and four TPUs are utilized.
Comparing LLM inference performance, we observe that \textit{Design A} achieves an average 28\% speedup over the baseline MXU configuration, while enjoying a remarkable 24.2$\times$ reduction in MXU energy.
Due to the higher peak performance, \textit{Design B} achieves 33\% throughput improvement compared to the baseline. 
The CIM-MXU also exhibits 6.34$\times$ energy reduction compared to the baseline systolic array MXUs. 

\subsection{Related Work}
\textbf{Modeling for LLM inference.}
Prior work LLMCompass \cite{zhang_hardware_2023} is a hardware evaluation framework tailored to LLM inference, providing hardware architecture templates and accurate performance and area evaluations. LLMCompass explored cost-effective hardware designs, highlighting that current acceleration hardware can be over-provisioned with wasted computation components. However, LLMCompass only focused on LLMs without considering the other mainstream generative models, i.e. Diffusion Models. At hardware level, it only adopts full digital implementation of acceleration hardware and does not explore other efficient computation paradigms, such as compute-in-memory in our work.

\textbf{CIM Simulators.}
Several CIM simulators have been proposed to evaluate the benefits of CIM for DNNs and Transformers.
For example, \cite{andrulis_cimloop_2024} introduces a CIM modeling framework, focusing on energy modeling, DNN model mapping, and cross-stack design explorations. \cite{zhu_mnsim_2023} presents a hierarchical CIM modeling structure that supports evaluating computing accuracy.
However, none of these prior works explore the use of CIMs as the key computational block in high-performance accelerator chips, such as TPUs.
To the best of our knowledge, our work is the first architectural modeling, analysis, and design exploration of a CIM-based TPU.

%% file: 6_conclusion.tex
\section{Conclusion}
In this work, we explore leveraging digital CIM technique in TPUs to enhance efficiency and performance for generative model inference.
We refer the baseline TPUv4i architecture and construct a CIM-based TPU model.
A CIM-MXU design is presented by organizing multiple CIM cores as systolic datapath to replace the vanilla digital MXU in TPU, achieving 9.43$\times$ energy and 2.02$\times$ area efficiency improvements.
We evaluate the inference breakdown of mainstream generative models, including LLMs and DiTs, to analyze CIM design benefits.
Furthermore, we explore architectural design choices for CIM-MXUs and present optimized TPU designs for LLMs and DiTs to enhance efficiency and performance. We also observed that the CIM design benefits can be scaled for multi-device TPUs.

\section*{Acknowledgement}
This work was supported in part by NSFC Grant No. 92464202.